# EC P-256: Successful Simple Power Analysis

Ievgen Kabin[1], Zoya Dyka[1], Dan Klann[1] and Peter Langendoerfer[1,2]
[1]*IHP – Leibniz-Institut für innovative Mikroelektronik*, Frankfurt (Oder), Germany
[2]*BTU Cottbus-Senftenberg*, Cottbus, Germany
{kabin, dyka, klann, langendoerfer}@ihp-microelectronics.com

*Abstract*—In this work we discuss the resistance of atomic pattern algorithms for elliptic curve point multiplication against simple side channel analysis attacks using our own implementation as an example. The idea of the atomicity principle is to make *kP* implementations resistant against simple side channel analysis attacks. One of the assumptions, on which the atomicity principle is based, is the indistinguishability of register operations, i.e. two write-to-register operations cannot be distinguished if their old and new data values are the same. But before the data can be stored to a register/block, this register/block has to be addressed for storing the data. Different registers/blocks have different addresses. In praxis, this different and key dependent addressing can be used to reveal the key, even by running simple SCA attacks. The key dependent addressing of registers/blocks allows to reveal the key and is an inherent feature of the binary *kP* algorithms. This means that the assumption, that addressing of different registers/blocks is an indistinguishable operation, may no longer be applied when realizing *kP* implementations, at least not for hardware implementations.

## I. INTRODUCTION

Elliptic Curve Cryptography (ECC) is nowadays the main mean for implementing different security services. Accelerated in hardware, cryptographic protocols can be applied also for resources constrained devices, for example in WSN and for IoT. ECs over prime and over extended binary finite fields are standardized by NIST in the USA [1] and by ENISA in Europe [2]. EC P-256 is also standardized for use in automotive [3].

The main operation in EC cryptographic protocols is the multiplication of an EC point *P* with a scalar *k*, denoted as *kP*. The security of the ECC is based on the secrecy of private keys, corresponding to Kerckhoff's principle [4]. The goal of attackers is to reveal the private key of an attacked person or device. In some EC protocols, for example, for authentication, the scalar *k* used in the *kP* calculation is the private key. In other protocols, for example, for signature generation [5]-[6], the scalar *k* is not the private key, but an attacker can easily calculate the private key if the scalar *k* is revealed.

*kP* is a complex operation that can be expressed as a sequence of finite field operations – multiplications, divisions, additions, subtractions as well as storing of intermediate results into registers and reading these data from registers/blocks. *kP* operations are executed on physical devices, i.e. a *kP* execution has its duration, energy consumption and EM emanation. These measurable physical parameters/effects are known as side channel effects. They depend on the implemented algorithm, technology in which the electrical circuit is implemented as well as on the input values of the algorithm, i.e. on the coordinates of an EC point *P* and on the scalar *k*. Analysing a measured power or a measured electromagnetic trace of a single *kP* execution with the goal to reveal the scalar *k* is known as horizontal side channel analysis (SCA) attacks. In many algorithms, the scalar *k* is processed bitwise, especially in hardware implementations of *kP* accelerators. Processing a key bit value '0' requires only an EC point doubling operation, but processing a key bit value '1' requires an EC point doubling and an EC point addition operation corresponding to the binary left-to-right and right-to-left *kP* algorithms that are the oldest, well-known and simplest EC point multiplication algorithms [7]. Point doubling and point addition are different operations and usually consist of different sequences of field operations. Consequently, their power profiles can be distinguished from each other. Attackers can use different signal processing methods, statistical or artificial intelligence methods for separating the power profiles of '0' bit vales from the '1' bit values. The designer's strategy to reduce the success of SCA attacks is to make the power profiles for processing all key bits indistinguishable from each other, i.e. independent of the processed key bit values. This strategy is known as the regularity principle if designers implement the same sequence of operation for the processing of each key bit. A simple way to make an algorithm regular is executing dummy operations, for example, a point doubling and a point addition can be calculated always for processing of each key bit value [8]. This increases the execution time and energy consumption significantly. Thus, the next idea was to make point doublings indistinguishable from point additions. If the power profile of a point doubling is the same as the one of a point addition, an attacker – theoretically – does not know, where a single point doubling for processing of a key bit '0' was performed and where a point doubling and point addition for processing of a key bit '1'. Different unified formulae for calculating point doublings and point additions using the same sequence of operations, realized by dummy field operations, are known. The number of dummy field operations was reduced by introducing the atomicity principle. Point doublings, as well as point additions, were represented as repeatable short sequences of field operations. In [9] the proposed short sequence – an atom – is the "multiplication, addition, negotiation, addition" ("M-A-N-A" atom), whereby a point doubling can be performed with 10 such atoms and a point addition with 16 such atoms. Other atoms were proposed in [10]. For example, 4 atoms consisting "M-N-A-M-N-A-A" are a point doubling and 6 of such atoms are a point addition. In [11] only 1 atom consisting of 8 multiplications, 6 additions, and 4 subtractions represents a point doubling and two such atoms represent a point addition. In [12] only a single atom consisting of 10 multiplications, 5 additions, and 5 subtractions represents a point doubling as well as a point addition.

The regularity strategy, unified point addition formulae, and atomicity principle are countermeasures against simple SCA attacks. But all these countermeasures are based on the assumption that power profiles of the processing of the same sequence of operations are indistinguishable.

In this paper we show on the example of the atomicity algorithm [12] implemented for an acceleration of cryptographic operations using the EC P-256 that the scalar $k$ can be fully revealed performing a simple SCA attack. The vulnerability of the atomicity algorithms is caused due to the key-dependent addressing of the registers and other blocks of the $kP$ design, i.e. our attack is a horizontal address bit attack.

The rest of this paper is structured as follows. In section II we explain the implementation details that are essential for understanding the nature of the address bit vulnerability. In section III we describe how we performed the automated simple power analysis attack and discuss its results. We demonstrate the vulnerability of atomic patterns to horizontal address bit SCA attacks using some operations in the atomic patterns as examples. The paper finishes with short conclusions.

## II. IMPLEMENTATION DETAILS

As we are focussing on the issue of the addressing of registers in this paper, we omitted all implementation details that are not essentially needed for understanding, what the address-bit leakage source is.

TABLE I. shows the operation sequence we implemented in our $kP$ design based on the algorithm proposed in section 4.1 in [12]. The original atomic patterns for EC point doublings and point additions [12] are, also given in TABLE I. .

We implemented the atomic patterns for point doublings as proposed in [12]. The left part of TABLE I. shows the sequence of the operations implemented in our design using our denotation for registers which is similar but not exactly the same as in [12].

We slightly modified the sequence for point additions with the goal to parallelize the operations to reduce the execution time and increase the resistance against SCA. We swapped the operations 2 and 3 as well as the operations 20 and 21 in the original atomic patterns, see TABLE II. .

After reordering these operations the pattern we used for point doubling as well as for point additions still are identical, i.e. this change is not the reason why we can distinguish between the two types of point operation processed.

TABLE I. ATOMIC PATTERNS FOR POINT DOUBLINGS AND POINT ADDITIONS FOR EC OVER GF(P)

| | in our design | | corresponding to [12] | |
|---|---|---|---|---|
| inputs: | $P=(X_1:X_2:X_3:Z_1)$ | $P=(X_1:X_2:X_3:Z_1:Z_2)$ $Q=(X:Y:1)$ | $P=(X:Y:Z:Z^2:Z^3)$ $Q=(X_q:Y_q:1)$ | $P=(X:Y:Z:Z^2)$ $I=1$ |
| Nr. | Doubling | Addition | Addition | Doubling |
| 1 | $R_0 \leftarrow X_3 \cdot X_3$ | $R_1 \leftarrow X \cdot Z_1$ | $R_1 \leftarrow X_q \cdot Z^2$ | $R_0 \leftarrow I \cdot Z^2$ |
| 2 | $R_2 \leftarrow X_2 + X_2$ | $R_2 \leftarrow X_2 + X_2$ | $R_1 \leftarrow R_1 - X$ | $R_1 \leftarrow X - R_0$ |
| 3 | $R_1 \leftarrow X_1 - R_0$ | $R_1 \leftarrow R_1 - X_1$ | $\star \leftarrow \star + \star$ | $R_2 \leftarrow Y + Y$ |
| 4 | $Z_1 \leftarrow X_2 \cdot R_2$ | $R_2 \leftarrow R_1 \cdot R_1$ | $R_2 \leftarrow R_1 \cdot R_1$ | $Z_2^2 \leftarrow Y \cdot R_2$ |
| 5 | $X_2 \leftarrow Z_1 + Z_1$ | $R_0 \leftarrow R_2 + R_2$ | $\star \leftarrow \star + \star$ | $Y_2 \leftarrow Z_2^2 + Z_2^2$ |
| 6 | $X_1 \leftarrow R_2 \cdot X_3$ | $R_3 \leftarrow X_1 \cdot R_2$ | $R_3 \leftarrow X \cdot R_2$ | $R_3 \leftarrow R_2 \cdot Z$ |
| 7 | $R_2 \leftarrow X_2 \cdot X_1$ | $R_0 \leftarrow Y \cdot Z_2$ | $R_0 \leftarrow Y_q \cdot Z^3$ | $R_2 \leftarrow Y_2 \cdot X$ |
| 8 | $X_1 \leftarrow X_1 + R_0$ | $Z_2 \leftarrow Z_2 + R_0$ | $\star \leftarrow \star + \star$ | $X_2 \leftarrow X + R_0$ |
| 9 | $R_0 \leftarrow R_1 \cdot X_1$ | $Z_2 \leftarrow R_1 \cdot R_2$ | $Z^3 \leftarrow R_1 \cdot R_2$ | $R_0 \leftarrow R_1 \cdot X_2$ |
| 10 | $R_1 \leftarrow Z_1 \cdot X_2$ | $R_2 \leftarrow X_3 \cdot R_1$ | $R_2 \leftarrow Z \cdot R_1$ | $R_1 \leftarrow Z_2^2 \cdot Y_2$ |
| 11 | $X_1 \leftarrow R_0 + R_0$ | $X_1 \leftarrow R_3 + R_3$ | $X_3 \leftarrow R_3 + R_3$ | $X_2 \leftarrow R_0 + R_0$ |
| 12 | $R_0 \leftarrow R_0 + X_1$ | $X_1 \leftarrow Z_2 + X_1$ | $X_3 \leftarrow Z^3 + X_3$ | $R_0 \leftarrow R_0 + X_2$ |
| 13 | $X_1 \leftarrow (R_0)^2$ | $Z_1 \leftarrow (X_1)^2$ | $Z_2^3 \leftarrow (R_2)^2$ | $X_2 \leftarrow (R_0)^2$ |
| 14 | $X_1 \leftarrow X_1 - R_2$ | $R_0 \leftarrow R_0 - X_2$ | $R_0 \leftarrow R_0 - Y$ | $X_2 \leftarrow X_2 - R_2$ |
| 15 | $Z_1 \leftarrow (R_3)^2$ | $R_1 \leftarrow (R_0)^2$ | $R_1 \leftarrow (R_0)^2$ | $Z_2^2 \leftarrow (R_3)^2$ |
| 16 | $X_1 \leftarrow X_1 - R_2$ | $X_1 \leftarrow R_1 - X_1$ | $X_3 \leftarrow R_1 - X_3$ | $X_2 \leftarrow X_2 - R_2$ |
| 17 | $R_2 \leftarrow R_2 - X_1$ | $R_1 \leftarrow R_3 - X_1$ | $R_1 \leftarrow R_3 - X_3$ | $R_2 \leftarrow R_2 - X_2$ |
| 18 | $Z_2 \leftarrow Z_1 \cdot R_3$ | $R_3 \leftarrow R_1 \cdot R_0$ | $R_3 \leftarrow R_1 \cdot R_0$ | $Z_2^3 \leftarrow Z_2^2 \cdot R_3$ |
| 19 | $X_2 \leftarrow R_0 \cdot R_2$ | $R_0 \leftarrow X_2 \cdot Z_2$ | $R_0 \leftarrow Y \cdot Z^3$ | $Y_2 \leftarrow R_0 \cdot R_2$ |
| 20 | $X_3 \leftarrow R_3$ | $X_3 \leftarrow R_2$ | $Y_3 \leftarrow R_3 - R_0$ | $Y_2 \leftarrow Y_2 - R_1$ |
| 21 | $X_2 \leftarrow X_2 - R_1$ | $X_2 \leftarrow R_3 - R_0$ | $Z_3 \leftarrow R_2$ | $Z_2 \leftarrow R_3$ |
| outputs: | $2P=(X_1:X_2:X_3:Z_1:Z_2)$ | $P+Q=(X_1:X_2:X_3:Z_1)$ | $P+Q=(X_3:Y_3:Z_3:Z_3^2)$ | $2P=(X_2:Y_2:Z_2:Z_2^2:Z_2^3)$ |

TABLE II. OUR MODIFICATION OF ATOMIC PATTERNS FOR POINT ADDITIONS

| operation in our design | operation in [14] |
|---|---|
| Nr. 2 | Nr. 3 |
| Nr. 3 | Nr. 2 |
| Nr. 20 | Nr. 21 |
| Nr. 21 | Nr. 20 |

We used the implemented atomic patterns for realizing the $kP$ operations in the binary double-and-add left-to-right algorithm for. The sequence of operations in TABLE I. consists of multiplications, additions and subtractions of elements of $GF(p)$ as well as write to register operations. Thus, our design consists of functional blocks for addition, subtraction and multiplication of $GF(p)$ elements for the EC P-256, as well as registers. The field multiplier needs 2 clock cycles for obtaining of new operands and only 9 additional clock cycles for the calculation of the field product. We achieved such a short time for calculating the product implementing the 4 segment Karatsuba multiplication formula [13]. The classical multiplication formula requires if the same segmentation of operands is applied (and, consequently, using the same partial multiplier) 16 clock cycles. Thus, the 4 segments Karatsuba multiplication method reduces the number of required partial products as well as the time and energy for their calculation of about $(16-9)/16 \cdot 100\% \approx 44\%$. Obtaining the new operands for the next multiplication can be done in parallel to the calculation of the last 2 partial products of the current multiplication. The addition and subtraction of the field elements requires 3 clock cycles: 2 clock cycles for obtaining the operands and 1 clock cycle for their processing. The sequence of the field operations as well as the storing of the data into/from blocks is managed by the block Controller. The communication between the functional blocks and the registers is realised by a multiplexer that we denote as bus.

Only one of blocks/registers during a clock cycle can write its output data to the bus. Controller determines this source block as well as the block(s) that receive the data from the bus. The addressing of the source and the destination blocks is a strong side channel leakage source, due to the fact that this addressing – corresponding to the *kP* algorithms – is key dependent.

In order to analyse our design we synthesized the design for the IHP 250 nm cell library SGB25V [14] using Synopsys Design Compiler Version K-2015.06-SP2 for the clock cycle period of 30 ns. We simulated a power trace of a *kP* execution using Synopsys PrimeTime (R) Version Q-2019.12-SP1. We set the time step between two simulated power samples to 0.01ns. Due all the parallelization we implemented our design needs 109 clock cycle for an EC point addition as well as for a doubling. Thus, a single atomic pattern consists of 32 700 samples.

### III. AUTOMATED SPA ATTACK

We performed a simple power analysis of the simulated power trace. The simulated power trace is a key dependent sequence of point doublings and point additions. The scalar *k* used in the *kP* execution for the power trace simulation is 255 bit long, contains 145 '1' and 110 '0'. Thus, the number of point doublings in this power trace is higher than the number of point additions: the simulated trace consists of 64% of atomic point doubling patterns and only 36% of atomic point additions. So trace consists of 255 doubling patterns and 145 addition patterns, i.e. 400 atomic patterns together. For each atomic pattern 32700 samples are recorded. This results into more than 13 million samples in the *kP* trace. We did not apply any compression of the trace, but due to the huge number of samples we automated the SPA. We calculated the "mean" atomic pattern and used it as a kind of threshold with the goal to distinguish doublings from additions. We compared each atomic pattern with the threshold sample-by-sample to make a decision about the "sort" of each atomic pattern. Thus, we obtained 32700 "key" candidates, one per sample of the atomic pattern. Each of the "key" candidates is a sequence of point doublings and additions. For evaluation of the attack success, we compared each "key" candidate sequence with the sequence of the patterns of the scalar *k*. For each "key" candidate we calculated its relative correctness, i.e. the relation of the number of the correctly revealed atomic patterns to the whole number of the patterns in the trace in per cent.

Fig. 1 shows the correctness of the obtained "key" candidates. The *x*-axis gives the index number of each "key" candidate as well as the index number of the clock cycle in the atomic pattern. Fig. 1 shows clearly that a lot of "key" candidates obtained are identical to the real sequence of the point operations in the attacked trace.

We are aware that the success of our attack is significantly reduced due to our selection of the threshold and due to the fact that the number of the point additions is not equal to the number of the point doublings, i.e. it is a kind of "worst case scenario" with respect to attack success. Despite this, we revealed the processed scalar *k* completely.

To demonstrate the fact, that the key dependent addressing is the reason of the successful SPA, we show the sequence of operations in the atomic patterns schematically, clock by clock, in a diagram overlapping the graph showing the correctness of the "key" candidates in Fig. 1.

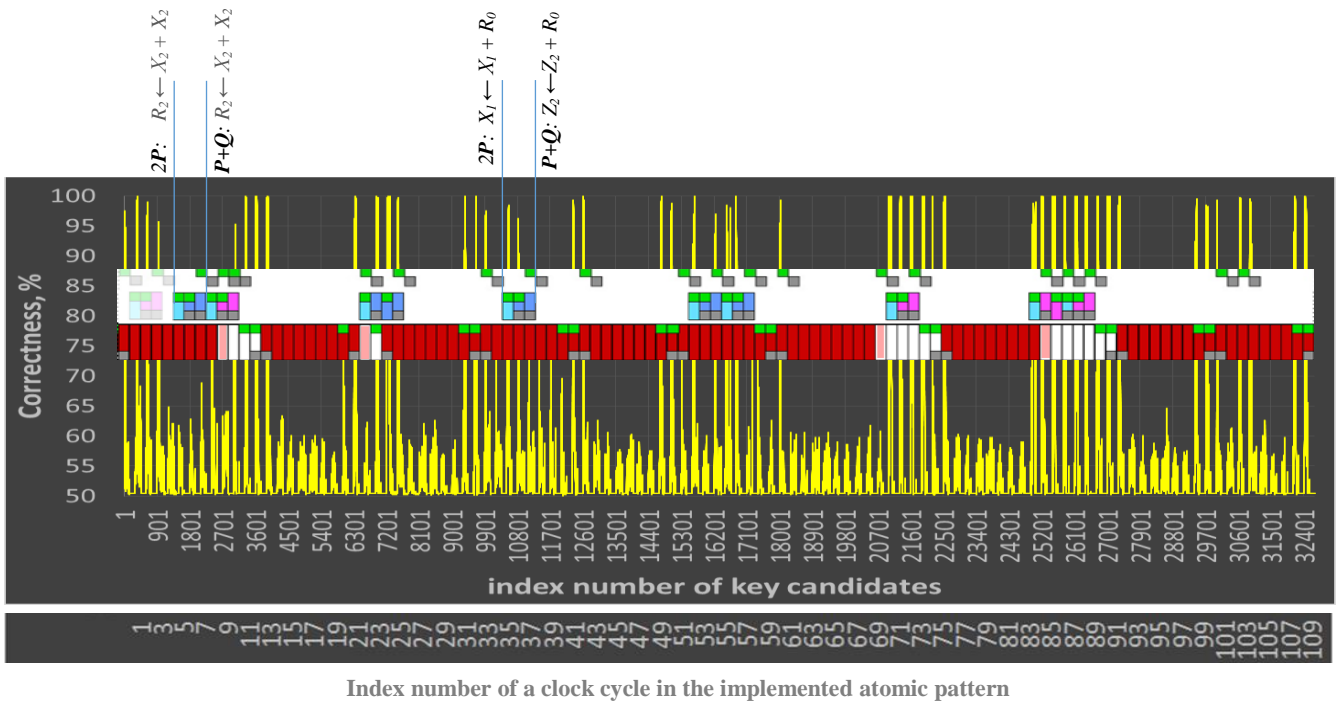

Fig. 1. Correctness of the "key" candidates..

The diagram showing the atomic patterns can be separated into 3 layers:

- The top layer shows small green and grey rectangles corresponding to the activity of the registers. Each green rectangle shows the addressing of a register for receiving a new input value. The register stores the data in the next clock cycle after the addressing; this is marked by the grey rectangles.

- The middle layer with the blue and magenta rectangles corresponds to the activity of the block for addition (blue) and subtraction (magenta) of the field elements. These operations require 3 clock cycles:
    - clock cycle 1: addressing of the block for receiving of its 1st operand (marked by a small green rectangle): the operand value is available on the bus
    - clock cycle 2: addressing of the block for receiving of its 2nd operand (marked by a small green rectangle): the operand value is available on the bus; the 1st operand already stored in its internal register (small grey rectangle) and also available. Thus, both operands can be processed.
    - clock cycle 3: storing the operation's result in the internal register (small grey rectangle): in this clock cycle the block can be addressed for writing the calculated value to the bus. This addressing is not shown in the pattern diagram.

- The bottom layer with red, light-red and white rectangles corresponds to the activity of the field multiplier. The product calculation requires 2+9+1=12 clock cycles:
    - clock cycle 1: addressing the multiplier for receiving its 1st operand (marked by a small green rectangle): the operand value is available on the bus
    - clock cycle 2: addressing the multiplier for receiving its 2nd operand (marked by a small green rectangle): the operand value is available on the bus; the 1st operand is already stored in its internal register (small grey rectangle).
    - clock cycle 3: the 2nd operand is stored in its 2nd internal register (small grey rectangle): the calculation of the partial products starts.
    - clock cycles 4-11: calculation of the partial products and accumulation of the result in the output register of the multiplier, including the field reduction (red rectangle).
    - clock cycle 12: the output register of the multiplier contains the field product; the multiplier can be addressed for writing the calculated value to the bus. This addressing is not shown in the pattern diagram. This clock cycle is marked in light-red if the multiplier does not calculate a partial product but consumes energy due to its first waiting clock cycle.

The next waiting cycles of the multiplier are marked by white rectangles. The multiplier waits, if the sequence of operations cannot be parallelized. For 6 of 10 multiplications the parallelization was possible: receiving a new multiplicand is done in parallel to the calculation of the 8th and 9th partial products. In addition addressing the multiplier for writing its output to the bus can be done in parallel to the calculation of the 1st partial product.

The atomic pattern starts with the 1st multiplication (see operation Nr.1 in TABLE I. ), specifically with its first of the nine partial multiplications. The comparison of the attack results with the atomic pattern diagram shows clearly, that the clock cycles corresponding to the addressing of different blocks (see green rectangles in all layers of the diagram) are SCA leakage sources. The energy consumed during a clock cycle depends on the address of the blocks in the current and the previous clock cycle. For example, at the end of the last field multiplication (see operation Nr. 19 in TABLE I. ) the multiplier is addressed to write its output to the bus in both atomic point operation patterns. But in case of a point doubling the register $X_2$ is addressed for receiving the new value and in case of a point addition it is another register – the register $R_0$ that has different address. This operation dependent addressing is performed in the 1st clock cycle shown in Fig. 1 and causes the high success rate of the attack, i.e. about 98% of the atomic patterns were correctly classified into doubling or addition even though the threshold for distinguishing between the operations is biased by an unbalanced number of the two patterns.

The vulnerability of atomicity patterns to the address bit attacks can be clearly demonstrated using the following example:

- 1st addition in the patterns (see operation Nr. 2 in TABLE I. ).
  In both point operation patterns completely the same operations are performed, including the used register and addressing aspects: $R_2 \leftarrow X_2 + X_2$. The correctness of all "key" candidates for the clock cycles 6-8 in the atomic pattern diagram (i.e. during this addition) is less than 68%.

- 3rd addition in the patterns (see operation Nr. 8 in TABLE I. ), i.e. we will compare the following two operations:
    - $X_1 \leftarrow X_1 + R_0$ for a point doubling;
    - $Z_2 \leftarrow Z_2 + R_0$ for a point addition.
  The first operand will be received from different registers and the sum will be saved also to different registers. The address of the register $R_0$ that is the source for the 2nd operand is the same. Due to the fact that not the address itself but the difference of the current and previous addresses causes the energy consumption that is specific for this difference, we expected a high success rate of the address bit attacks because of:
- o in case of a point doubling it is the difference of the addressing of $X_1$ in previous and $R_0$ in the current clock cycle;
- o in case of a point addition it is the difference of the addressing of $Z_2$ in previous and $R_0$ in the current clock cycle.

The attack results confirm our explanation: the highest correctness of the "key" candidates is about of 99, 97 and 78 per cent for the 1st, 2nd and 3rd clock cycle of this addition, respectively (see clock cycles 35, 36 and 37 in the atomic pattern diagram).

We selected these two additions for the comparison of their vulnerability to address bit attacks due the fact that parallel to both additions only partial products in the field multiplier are calculated, i.e. there is no other addressing activity. This fact makes the comparison fair.

## IV. CONCLUSION

In this paper we discussed the vulnerability of atomic patterns to simple SCA attacks. We implemented a binary double-and-add left-to-right $kP$ algorithm for the NIST EC P-256. We simulated a power trace of a $kP$ execution and performed an automated simple power analysis attack without any trace compression. We were able to classify the attacked trace fully correct into doubling and addition parts. This allows to reveal the used scalar $k$ completely. Due to the fact that we attacked an uncompressed trace, we obtained 32700 key candidates. The number of key candidates revealed with the correctness of 100% is 348.

The vulnerability of the atomicity algorithms is caused due to the key-dependent addressing of the registers and other blocks in algorithms for calculating the elliptic curve point multiplication, i.e. it is an inherent feature of the algorithms. In our experiments we attacked a single $kP$ execution, i.e. we performed a simple horizontal address bit SCA attack.

The addressing of registers was already earlier identified as a leakage that allows to successfully distinguish which key bit '1' or '0' was processed when. E.g. Itoh et al. presented a successful vertical, address bit differential power analysis (DPA) attacks against ECC, already in 2002 in [15]. Also the Montgomery ladder using Lopez-Dahab projective coordinates for EC over $GF(2^n)$ is known to be vulnerable to horizontal, i.e. single-trace, address bit DPA [16].

Thus, due to results published in [15]-[16] and here, the common assumption that addressing of different registers/blocks is an indistinguishable operation has to be revised, at least for hardware implementations. .

ACKNOWLEDGMENT

The work presented here was supported by the Federal Ministry of Education and Research (BMBF) of Germany under grant number 03ZZ0527A.